\documentclass{article}
\usepackage{graphicx}
\usepackage{subcaption}
\usepackage{amssymb}

\title{Radiation-mediated shocks: kinetic 
processes and transition to 
collisionless shocks}
\author{Evgeny Derishev\\
\small\itshape Institute of Applied Physics, Nizhny Novgorod, Russia}
\date{December 2018}

\begin{document}

\maketitle

\begin{abstract}
We analyze the main features of radiation-mediated shocks at arbitrary shock velocities, both non-relativistic and relativistic. We describe two mechanisms, which may lead to formation of a sharp viscous subshock within otherwise smooth velocity profile at the shock front, even if the radiation pressure in the upstream is overwhelmingly large. These mechanisms are specific to sub-relativistic and relativistic radiation-mediated shocks and set high-velocity shocks apart from their non-relativistic counterparts, which do not develop a subshock if the radiation pressure is high enough. We briefly discuss implications of this finding.
\end{abstract}

\section{Introduction}\label{introduction}

Astrophysical shocks usually propagate in collisionless plasma, so that the viscosity and hence the shock front width are determined by plasma turbulence, which couples various particle species on scales larger than the plasma skin depth
\begin{equation}
	l_\mathrm{p} = \left(\frac{m_\mathrm{i}c^2}{4\pi (Z e)^2 N_\mathrm{i}}\right)^{1/2}.
\end{equation}
Here $Ze$ and $m_\mathrm{i}$ are charge and mass of ions (usually protons) and $N_\mathrm{i}$ their number density. Coupling of radiation to the plasma occurs on much longer scales, determined by photons' mean free path $\lambda$, which is typically many orders of magnitude larger than the plasma skin depth. On scales much larger than $\lambda$ one can still consider any shock as a discontinuity, where the jump conditions are set by conservation of energy and momentum fluxes with photons' contribution to pressure and enthalpy taken into account. On scales between $\lambda$ and $l_\mathrm{p}$ the shock front is resolved into a long precursor of length $\sim \lambda$, where photons escaping from downstream heat and accelerate the plasma, and a subshock, where density and velocity of the plasma undergo a jump, which conserves energy and momentum fluxes for the plasma, excluding contribution from photons. The subshock has a width of the order of $l_\mathrm{p}$ and the photons are not coupled to the plasma on this scale.

There are at least three situations in astrophysics, where one finds radiation densities so large, that contribution of radiation to the total pressure is dominant and the shock front structure is mostly determined by coupling between radiation and plasma. Such shocks, usually termed radiation-mediated shocks (RMSs), can be found inside exploding supernovae, in accretion flows hitting the surface of neutron stars and white dwarfs, and in gamma-ray burst sources, both inside their jets and in stellar envelope during the jet breakout. 
There are analytic solutions for RMSs, obtained using diffusion approximation for the radiation and therefore valid for the case where the shock speed $U_\mathrm{sh}$ is much less than the speed of light \cite{Marshak,ZeldovichRaizer}. With these solutions, it was shown that if the ratio of radiation pressure to gas pressure, $P_\mathrm{r}/P_\mathrm{g}$ in the downstream exceeds critical value $\simeq 4.4$, then the subshock disappears and the transition from the upstream to the downstream becomes smooth on the scale $\sim l_\mathrm{p}$; in the book \cite{ZeldovichRaizer} first notion of this fact is attributed to S.Z. Belen'kii (1950).

There is a common agreement that in the three cases listed above, the shocks have radiation pressure above the critical value and, consequently, a smooth velocity profile at the shock front. In turn, the smooth velocity profile means that coupling between various plasma particle species can be maintained by plasma turbulence at a very small level, which would have no consequences apart from ensuring that plasma particles behave as a single fluid.
This implies a relatively simplified description of RMSs in the case of large photon-to-electron ratio, which includes processes of plasma-radiation coupling but excludes some processes essential for collisionless shocks, most notably -- particle acceleration. For this reason, recent papers are focused on details of photon interaction with plasma, such as photon transport at arbitrary shock velocity, photon generation and absorption, creation and annihilation of electron-positron pairs (e.g., \cite{BudnikWaxman, Blinnikov, Beloborodov, ItoLevinson, LundmanBeloborodov}).

In this paper, we argue that sub-relativistic or relativistic RMSs with photon-to-electron number ratio in the upstream well above the critical value may still develop a subshock. The conditions favoring re-appearance of the subshock are outside the range of validity for existing analytic solutions, so that the problem can only be studied numerically or via qualitative estimations. Here we use both approaches. 

The paper is organized as follows. In section \ref{code} we describe the features and capabilities of the code used to model radiation-mediated shocks. Section \ref{review} gives a summary of essential features of RMSs. In the next two sections, \ref{ionheating} and \ref{pairsubshock}, we discuss two mechanisms, which may lead to formation of strongly turbulent (viscous) subshock in fast RMSs. In the last section we summarize the results and briefly discuss their implications.

\section{The code for numerical simulations}\label{code}

The code, which was employed for numerical modeling of RMSs, uses a combination of Particle-in-Cell method and Monte-Carlo method. At start, the particles are generated with given momentum distributions and each of them is assigned to a cell according to it's coordinates. 
At each step and for each cell the code performs collisions between various particles contained within the same cell. The probability of collision for each pair of particles is calculated using the total interaction cross-section and if the collision takes place, then the scattering angle is randomly chosen according to the differential cross-section. Velocities of particles do not change between collisions, so at each step particle coordinates change according to rectilinear motion law. Finally, the particles are moved to adjacent cells if their new coordinates are outside the cell's borders.

The code performs electron-photon and positron-photon collisions, as well as two-photon electron-positron pair production and electron-positron annihilation into two photons. For these interactions the exact differential QED cross-sections for unpolarized particles are used. The code also mimics strong coupling between electrons/positrons and ions by performing electron-ion and positron-ion collisions with artificially large and isotropic differential cross-section. 

Simulations used in this paper do not include processes of true emission and absorption. Also, the ion-to-electron mass ratio was set to 50 to avoid waste of computational resources -- the realistic large mass ratio would require proportionally larger number of collisions between electrons/positrons and ions to couple them into a single fluid.

\section{Main features of radiation-mediated shocks}\label{review}

Consider photon diffusion approximation used in analytic solutions for non-relativistic RMSs. In a stationary case, photons diffusing into the upstream form a precursor with exponentially decreasing number density of photons, as follows from stationary diffusion equation
\begin{equation}
	N_\mathrm{ph} U_\mathrm{sh} = D \frac{\partial N_\mathrm{ph}}{\partial x} \qquad \Rightarrow \qquad
N_\mathrm{ph} \propto \exp\left( \frac{U_\mathrm{sh}}{D} x \right),
\end{equation}
where the diffusion coefficient is $D=\lambda c/3$. In the typical case, where the photons interact with plasma mostly via Thomson scattering, $\lambda=1/(\sigma_\mathrm{T} N_\mathrm{e})$ and $D=c/(3\sigma_\mathrm{T} N_\mathrm{e})$; $N_\mathrm{e}$ is the total number density of electrons and positrons. Therefore, the shock front width is 
\begin{equation}
L_\mathrm{sh} \simeq D/U_\mathrm{sh} \qquad \mathrm{(non-relativistic\; case)}
\end{equation}
and the shock-front optical depth is $\tau_\mathrm{sh} \simeq c/(3 U_\mathrm{sh})$. Apparently, this result is valid only if $U_\mathrm{sh}\ll c/3$. For a fast shock with $U_\mathrm{sh}\gtrsim c/3$ one has to solve kinetic equation for photons, and the shock-front optical depth in this case is $\tau_\mathrm{sh} \simeq 1$.
A convenient expression, which interpolates between non-relativistic and relativistic cases is
\begin{equation}\label{shockwidth}
L_\mathrm{sh} \simeq D/U_\mathrm{sh} + \lambda\, .
\end{equation}

\begin{figure}
    \includegraphics[width=0.87\textwidth]{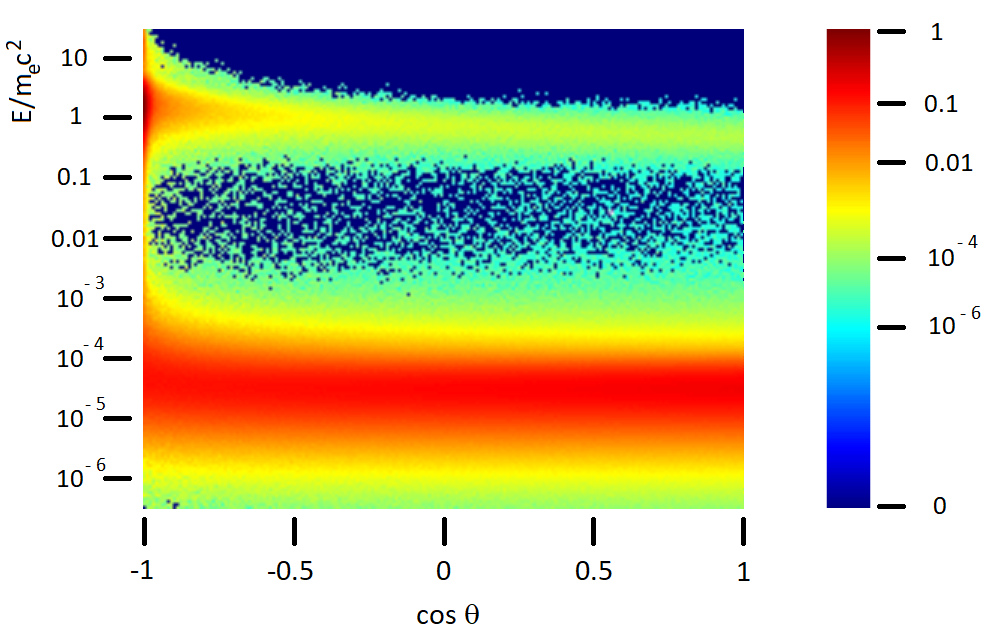}
		\caption{Distribution of photons in the plasma comoving frame taken in the upstream at optical depth $\simeq 5$ from the shock front. 
		Color denotes number of protons per logarithmic energy interval,
		$\theta$ is photon propagation angle relative to the shock normal. Shock speed is 0.95 $c$.}
		\label{PhotonDistribution}
\end{figure}

Figure~\ref{PhotonDistribution} shows photon distribution over energies and propagation angles taken from numerical simulation of a fast shock. It clearly demonstrates that photon distribution in the upstream has two components: nearly isotropic low-energy component is composed of photons advected by the flow from the far upstream. The second, high-energy, component includes photons escaping from the downstream. This component is highly anisotropic, which certainly renders diffusion approximation invalid.

\begin{figure}
    \includegraphics[width=0.87\textwidth]{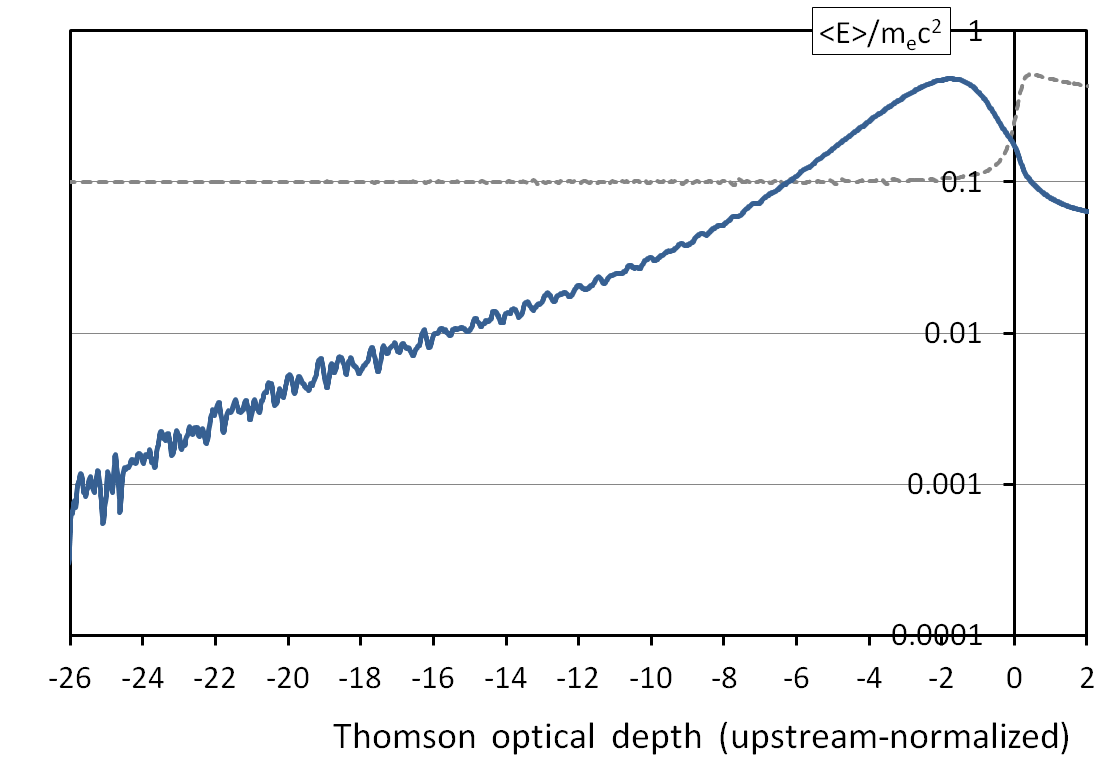}
		\caption{Average comoving-frame energy of ions. Solid line -- average energy normalized in units $m_e c^2$,
dashed line -- ion density (arbitrary units). Shock speed is 0.95 $c$.}
\label{temperature_profile}
\end{figure}

The photons escaping from the downstream heat the upstream plasma, forming exponentially rising temperature profile in the shock precursor. As can be seen from Fig.~\ref{temperature_profile}, the temperature peaks at some distance from the shock front, reaching values comparable to the downstream temperature.

Although diffusion approximation for radiation breaks in high-velocity RMSs, leaving us without analytic solution, it is still possible to show that such shocks also can organize themselves into smooth transition between the upstream and the downstream without forming sharp viscous subshock. Following the approach of \cite{DerishevPiran}, we write the flux conservation equations, 
\begin{equation}
w_1 \beta_1^2 \Gamma_1^2 + p_1 = w_2 \beta_2^2 \Gamma_2^2 + p_2 + S_\mathrm{mom}
\end{equation}
for momentum flux and 
\begin{equation}
	w_1 \beta_1 \Gamma_1^2 = w_2 \beta_2 \Gamma_2^2 - S_\mathrm{en}
\end{equation}
for energy flux, with two extra terms, $S_\mathrm{mom}$ and $S_\mathrm{en}$, which take into account difference between bulk velocities of radiation and plasma. Both terms vanish far in the upstream and far in the downstream. Here $\beta$ and $\Gamma$ are the bulk velocity of plasma (in units of speed of light) and bulk Lorentz factor of plasma, respectively, $p$ the total (plasma + radiation) pressure and $w$ the total specific enthalpy. It is convenient to normalize these extra terms: 
$S_{en} = a\, w_2 \beta_2 \Gamma_2^2$ and $S_{mom} = b\, S_{en}$.

If there is a steady state 1D solution, then instead of equating incoming and outgoing fluxes on both sides of a surface located at the shock front, one can equate them at opposite sides of two separated surfaces.
Putting the first surface into the transition region and the second far in the upstream, one obtains two solutions, corresponding to the plasma velocities upstream and downstream of the viscous subshock. The difference between the solutions becomes smaller as the feedback parameter $f=4(1-a)\beta_1 - 4ab$ increases, and goes to zero at the critical value $f_{cr}=\sqrt{3}/2$.
Existence of the critical value for the feedback parameter means that for high enough ratio of radiation pressure to the gas pressure the shock front organizes itself into a smooth structure without a subshock. Unfortunately, in absence of analytic solution it is not straightforward to translate the critical value of $f$ into critical ratio of radiation pressure to gas pressure, as it is done for non-relativistic RMSs.

\section{Runaway ion heating in the upstream}\label{ionheating}

Passing through the shock front, the ions decelerate in one of two ways. Coulomb collisions ensure coupling between ions and electrons (positrons) at timescales larger than
\begin{equation}
	t_\mathrm{c} \simeq \frac{m_\mathrm{i}}{m_\mathrm{e}} \frac{1}{\sigma_\mathrm{c} N_\mathrm{e} V_\mathrm{e}}, 
\end{equation}
where
\begin{equation}
 \sigma_\mathrm{c} = \frac{Z^2 r_\mathrm{e}^2\, \Lambda_\mathrm{c}}{\left(V_\mathrm{e}/c\right)^4}
\end{equation}
is the Coulomb collision cross-section, $\Lambda_\mathrm{c}$ the Coulomb logarithm, $r_\mathrm{e}$ the classical electron radius, and the fluid-frame electron velocity $V_\mathrm{e}$ can be estimated from the downstream temperature $T_\mathrm{d}$.
The actual timescale for ion deceleration is set by the shock front width:
\begin{equation}
	t_\mathrm{sh} \simeq \frac{L_\mathrm{sh}}{U_\mathrm{sh}}.
\end{equation}
Coulomb collisions cannot couple electrons and ions if $t_\mathrm{sh} \lesssim t_\mathrm{c}$, i.e., when
\begin{equation}
1 \lesssim \frac{t_\mathrm{c}}{t_\mathrm{sh}}  \simeq 
\frac{m_\mathrm{i}}{m_\mathrm{e}}\; \frac{\sigma_\mathrm{T}}{\sigma_\mathrm{c}}\; 
\frac{U_\mathrm{sh}}{\tau_\mathrm{sh} V_\mathrm{e}} \sim 
\frac{8\, m_\mathrm{i}}{\Lambda_\mathrm{c} m_\mathrm{e}}\; \beta_\mathrm{sh}^2
\left(\frac{T_\mathrm{d}}{m_\mathrm{e} c^2}\right)^{3/2}
\end{equation}
For a sub-relativistic shock, the Coulomb coupling is efficient only at temperatures $T_\mathrm{d} \lesssim 5$~keV.

At high temperatures, when the Coulomb collisions cannot couple electrons and ions, the work of ion deceleration is done by the electric field induced by charge separation. This field must extract all kinetic energy from ions, so that difference of potentials across the shock front is 
\begin{equation}
\Delta \phi = \frac{m_\mathrm{i} c^2 (\Gamma_\mathrm{sh}-1)}{Ze} \equiv E_0\, L_\mathrm{sh}\, .
\end{equation}
In the non-relativistic case, typical electric field strength at the shock front is 
\begin{equation}\label{field_estimate}
E \sim E_0 = \frac{\beta_\mathrm{sh}^3}{2}\, \frac{m_\mathrm{i} c^2}{Ze}\, \sigma_\mathrm{T} N_\mathrm{e}\, .
\end{equation}

\begin{figure}
     		\includegraphics[width=1.0\textwidth]{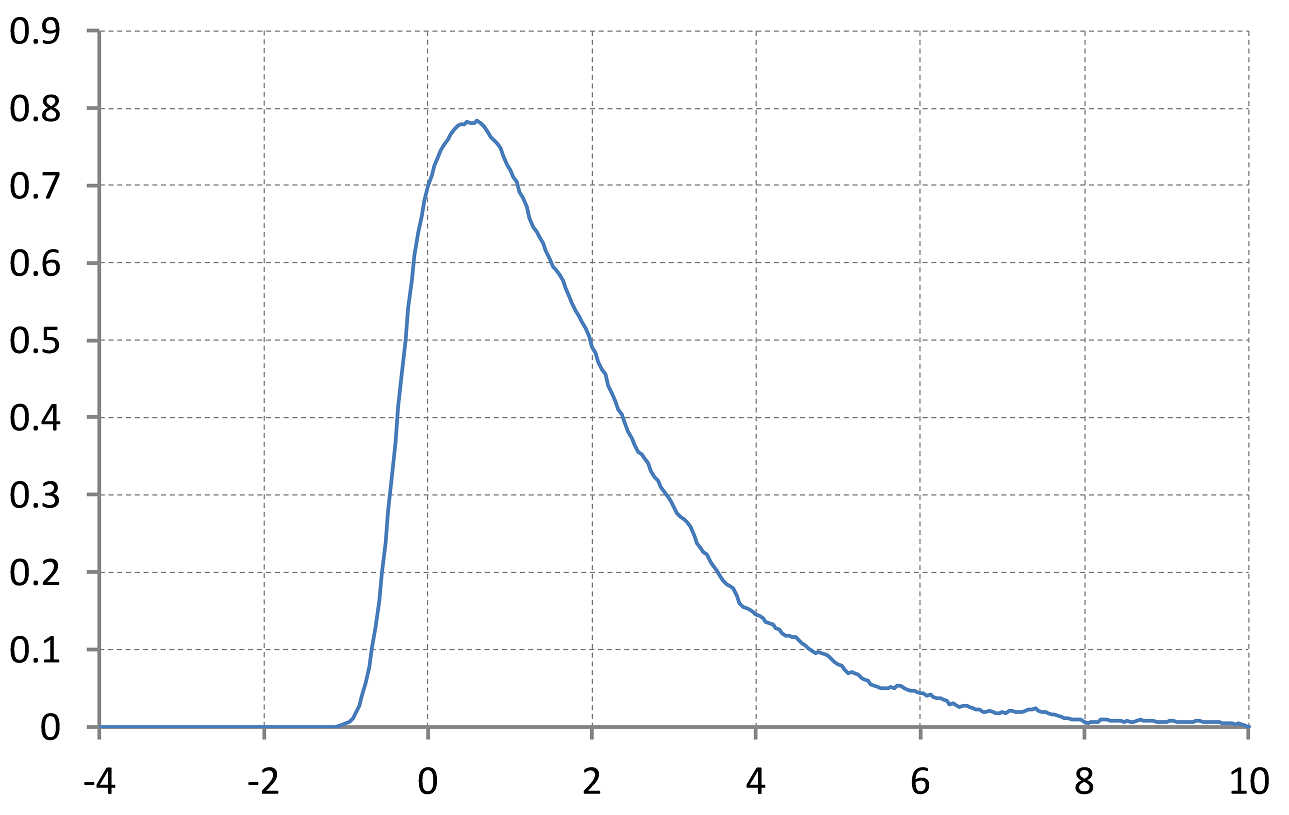}
\caption{Electric field at the shock.
Numerical simulation, shock speed $U_\mathrm{sh} = 0.6\, c$.
Vertical axis: local electric field in units $E_0$ (see Eq.~\ref{field_estimate}).
Horizontal axis: Thomson optical depth $\tau = x \sigma_\mathrm{T} N_\mathrm{e}$.
Origin of x-axis is at the point where $N_\mathrm{e} = 2 N_\mathrm{e}^{(u)}$.
}
\label{electric_field}
\end{figure}

One can calculate the decelerating electric field directly from the results of numerical simulations, assuming that the comoving-frame acceleration of ions, 
$c^2 d\Gamma/dx$ (here $\Gamma$ is the bulk Lorentz factor), is due to the electric field, so that $E = m_\mathrm{i} c^2 (d\Gamma/dx)/(Ze)$. An example of such calculation in presented in Fig.~\ref{electric_field}. It demonstrates that the simple analytic estimate (Eq.~\ref{field_estimate}) is fairly good.

Interacting with the electric field at the shock front, ions with different charge-to-mass ratio (typically, protons and $^4\mathrm{He}$ nuclei) move with different acceleration. This is equivalent to presence of effective electromotive force in the center-of-inertia frame of all ions, which can be estimated as
\begin{equation}
F_\mathrm{emf} \sim e E_0\, .
\end{equation}
The corresponding current density is
\begin{equation}
j \sim \frac{e^2 N_\mathrm{i}}{m_\mathrm{i} \nu_\mathrm{c}} E_0\, ,
\end{equation}
where $\nu_\mathrm{c}$ is the ion collision rate.
This current causes ohmic heating, so that the ion temperature increases at the rate 
\begin{equation}
\dot{T}_\mathrm{i} \sim \frac{j E_0}{N_\mathrm{i}} \sim \frac{e^2}{m_\mathrm{i} \nu_\mathrm{c}} E_0^2
\end{equation}
Solution of the above equation, $\dot{T}_i \propto T_i^{3/2}$, describes explosive rise of temperature, which tends to infinity at time $t_\mathrm{inf} = 2T_\mathrm{i}/\dot{T}_\mathrm{i}$. Thus, runaway ion heating occurs when
\begin{equation}
T_i \gtrsim \frac{\Lambda_c^2}{\beta_\mathrm{sh}^8}\left(\frac{m_e}{m_i}\right)^3 m_e c^2  \quad \Rightarrow \quad \beta_\mathrm{sh} \gtrsim 0.1\, c \, .
\end{equation}

Heating of ions stops when their beams become strong enough to excite fast-growing plasma instabilities, such as ion-acoustic instability. Although description of this stage poses serious difficulties, one may speculate that the outcome will be in formation of a layer of strong plasma turbulence, which would lead to re-appearance of the viscous subshock.

\section{Influence of pair creation}\label{pairsubshock}

With increasing upstream energy per photon, the downstream eventually becomes sufficiently hot to make two-photon pair production possible. In absence of true photon  absorption, both photons and electrons/positrons in equilibrium have relativistic Boltzmann distributions and their pressure ratio equals to the number density ratio. Neglecting upstream electron particle flux compared to that of photons, one finds that
\begin{equation}
\displaystyle
	\frac{P_\mathrm{r}}{P_\mathrm{g}} = \frac{N_\mathrm{ph}}{N_\mathrm{e}} 
	= \exp{\left(m_ec^2/T\right)} \times
	\frac{\int{p^2\exp{\left(-pc/T\right)}}\,\mathrm{d}p}{2\int{p^2\exp{\left(-\sqrt{m_e^2c^2+p^2}c/T\right)}\,\mathrm{d}p}}.
\end{equation}
At downstream temperatures exceeding $T_\mathrm{cr} \simeq 0.23 m_e c^2$, the above ratio exceed the critical value $\simeq 4.4$ (see Sect.~\ref{introduction}), suggesting formation of the viscous subshock. (Note, that the critical pressure ratio was obtained under certain assumptions for non-relativistic shocks. It can only serve as a proxy for sub-relativistic and relativistic shocks.) The critical temperature corresponds to critical upstream energy per photon
\begin{equation}
E_\mathrm{cr} = \frac{\left(\Gamma_\mathrm{sh}-1\right) m_\mathrm{i} c^2}{(N_\mathrm{ph}/N_\mathrm{e})Z} \simeq 1.1\, m_e c^2\,. 
\end{equation}

\begin{figure}
\centering
\begin{subfigure}[b]{0.49\textwidth}
   \includegraphics[width=\linewidth]{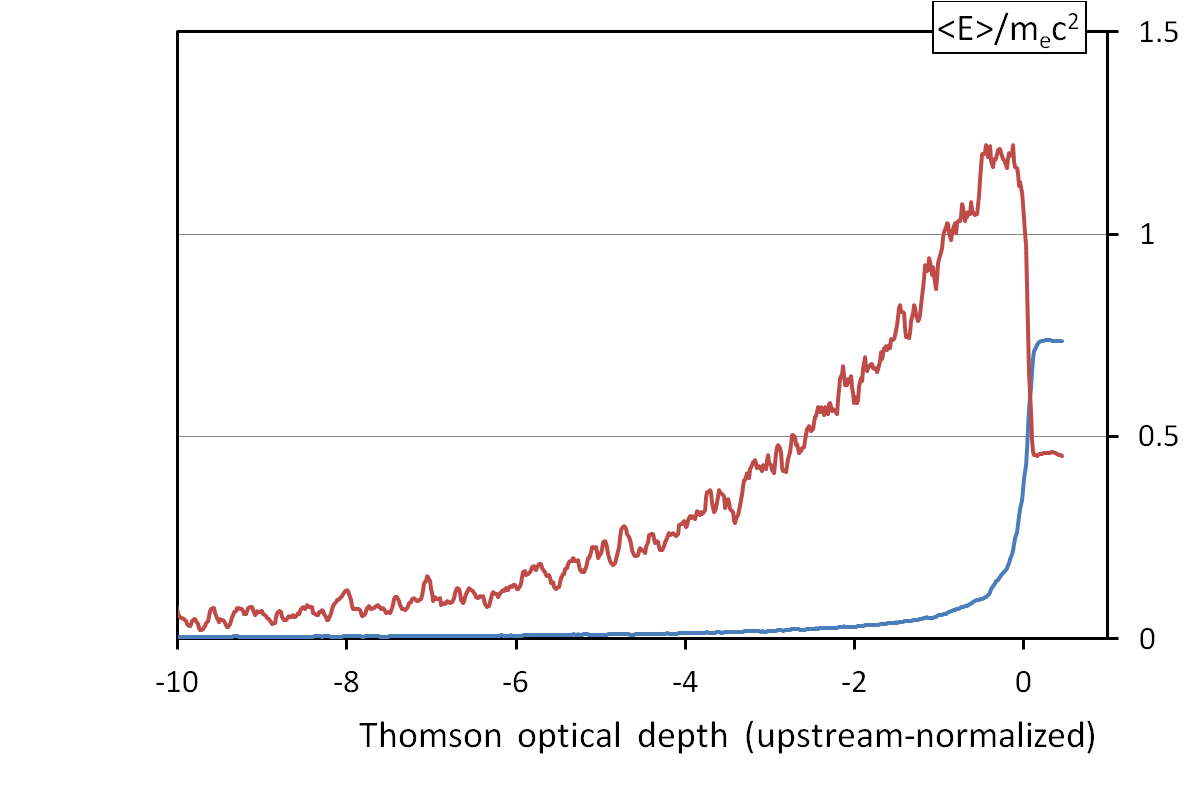}
\end{subfigure}
\begin{subfigure}[b]{0.49\textwidth}
    \includegraphics[width=\linewidth]{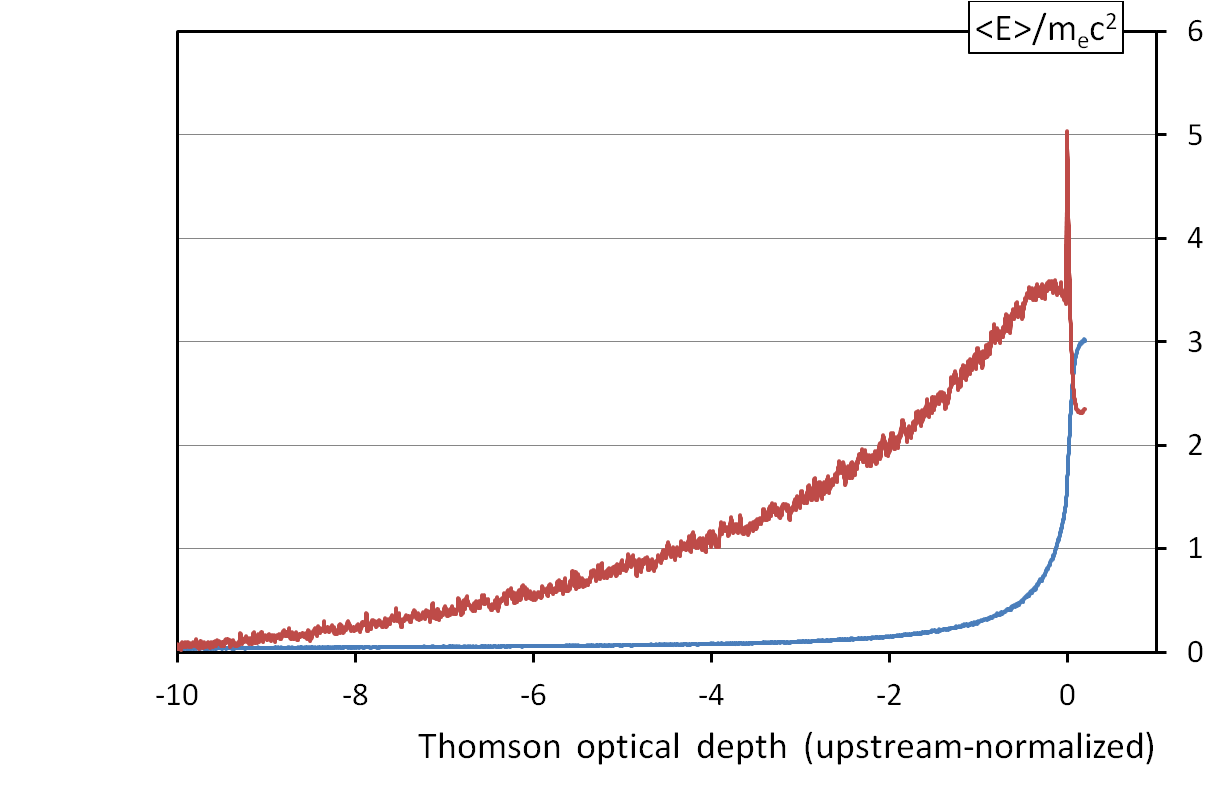}
\end{subfigure}

\begin{subfigure}[t]{0.49\textwidth}
    \includegraphics[width=\linewidth]{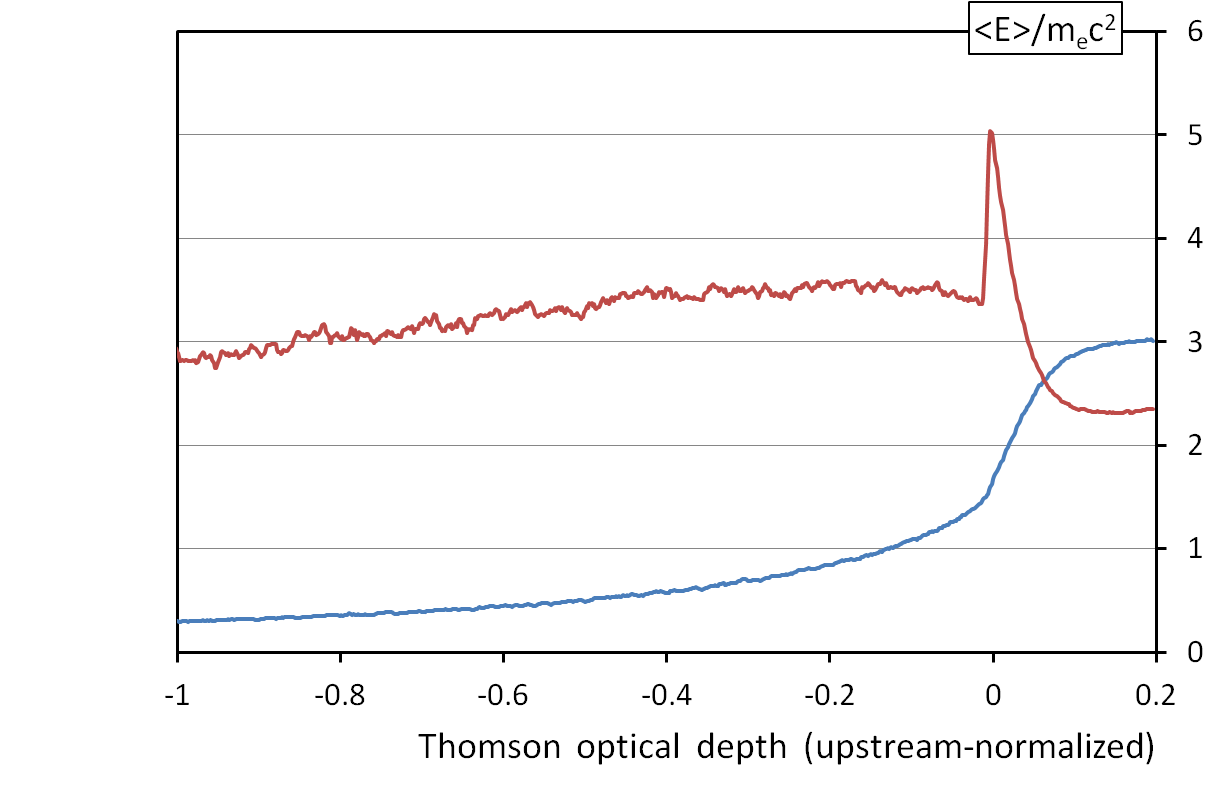}
\end{subfigure}
		\caption{ Average comoving-frame energy of ions as function of optical depth for simulations with
		ion-to-electron mass ratio 50, shock speed $0.95 c$, upstream temperature $10^{-3} m_e c^2$.
		Top left panel: upstream photon-to-electron number ratio is 100 (upstream energy per photon is $\simeq 1.63\, m_e c^2$).
		Top right panel: upstream photon-to-electron number ratio is 25 (upstream energy per photon is $\simeq 6.53\, m_e c^2$).
		Lower panel: blow-up of the subshock region from the top right panel.
		Horizontal axis: upstream-normalized Thomson optical depth $\tau = x \sigma_\mathrm{T} N_\mathrm{e}^{(u)}$.
		The origin of x-axis is at the point where the density is $N_\mathrm{e} = 2 N_\mathrm{e}^{(u)}$.}
		\label{subshock}
\end{figure}

Results of numerical simulations confirm formation of the subshock at large upstream energy per photon. The subshock unequivocally shows up as a sharp rise of ion temperature. Figure~\ref{subshock} demonstrates increase of subshock amplitude with increasing upstream energy per photon.

\section{Summary and implications}

Standard treatment of RMSs portrays them as objects with relatively 
simple physics, but without universal analytic solution. In this paper we demonstrate that physics of RMSs is richer and must include plasma kinetic processes. 
These processes hinder formation of smooth shock front in sub-relativistic and relativistic RMSs. 

There are two mechanisms, which may cause re-appearance of the viscous subshock even in the case where the radiation pressure is very large compared to the gas pressure. First mechanism is related to formation of counter-propagating ion beams as ions with different charge-to-mass ratio interact with charge-separation electric field at the shock front. This mechanism becomes efficient at $U_\mathrm{sh} > 0.1 c$ and results in excitation of strong plasma turbulence, most likely via ion-acoustic instability, and subsequently to formation of the viscous subshock. 
Accretion shocks in magnetized neutron stars and white dwarfs are prone to the effect of ion heating to even greater extent. Due to cyclotron resonance, the 
effective electron-photon cross-section can be 4 - 6 orders of magnitude larger, that means the ion heating withing the shock front would be 8 - 12 orders of magnitude faster.

The second mechanism for re-appearance of the viscous subshock is decreasing the downstream ratio of radiation pressure to the gas pressure below the critical value via massive production of electron-positron pairs. This effect inevitably takes place at sufficiently high downstream temperatures ($T_\mathrm{d} \gtrsim 0.23 m_\mathrm{e} c^2$), no matter how large was the radiation pressure dominance in the upstream. Fast shocks must be exceptionally efficient in producing photons to lower their downstream temperature to avoid formation of the viscous subshock in this way.
It should be noted that copious pair creation at the shock front does not interfere with the first mechanism, but rather facilitates it: the strength of Colulomb coupling between electrons/positrons and ions increases proportionally to the pair density, whereas the rate of ion heating is proportional to the square of the charge-separation electric field, i.e., to the square of pair density.

Formation of the viscous subshock is not only an interesting detail of shock front structure, it may qualitatively change the appearance of RMSs. For example, sufficiently strong subshocks may be capable of particle acceleration. The latter, if happens in a dense medium, may result in production of neutrinos via inelastic collisions of hadrons. There are good chances to observe neutrino emission from RMSs at their breakout in gamma-ray burst sources or even from the shocks inside exploding supernovae.

\section{Acknowledgement}

This work was supported by the Russian Science Foundation grant No 16-12-10528 (research on plasma kinetic instabilities at the shock front) and the RFBR grant No. 17-02-00525A (analysis and numeric simulations of pair production effect in RMSs).

\newpage
\bibliographystyle{unsrt}
\bibliography{references} 

\end{document}